\documentclass[aps,pra,twocolumn,showpacs,
floatfix]{revtex4-2}

\usepackage[utf8]{inputenc} 
\usepackage{amssymb,amsmath}
\usepackage{graphicx}
\usepackage{braket}
\usepackage{siunitx}
\usepackage{xcolor}
\usepackage[normalem]{ulem}
\usepackage{bm}
\usepackage{verbatim}

\def\Rb87{^{87}\mathrm{Rb}}                     
\def\ex{\mathbf{e}_x}  
\def\ey{\mathbf{e}_y}  
\def\ez{\mathbf{e}_z}  

\def\kr{k_0}
\def\Er{E_0}

\newcommand{\M}{\ensuremath{\hat{M}({\bf k}_\perp)}} 
\newcommand{\Mt}{\ensuremath{\hat{M}^\dagger({\bf k}_\perp)}} 

\begin{document}

\title{
Quantum Back-action Limits in Dispersively Measured Bose-Einstein Condensates
}

\author{Emine~Altunta\c{s}}
\email{altuntas@umd.edu}
\affiliation{National Institute of Standards and Technology, Gaithersburg, MD, 20899, USA.}
\affiliation{Joint Quantum Institute, University of Maryland, College Park, MD, 20742, USA.}
\author{I.~B.~Spielman}
\email{ian.spielman@nist.gov}
\affiliation{National Institute of Standards and Technology, Gaithersburg, MD, 20899, USA.}
\affiliation{Joint Quantum Institute, University of Maryland, College Park, MD, 20742, USA.}
\homepage{http://ultracold.jqi.umd.edu}
\date{\today}


\begin{abstract}
A fundamental tenet of quantum mechanics is that measurements change a system's wavefunction to that most consistent with the measurement outcome, even if no observer is present.
Weak measurements produce only limited information about the system, and as a result only minimally change the system's state.
Here, we theoretically and experimentally characterize quantum back-action in atomic Bose-Einstein condensates interacting with a far-from resonant laser beam.
We theoretically describe this process using a quantum trajectories approach where the environment measures the scattered light and present a measurement model based on an ideal photodetection mechanism. 
We experimentally quantify the resulting wavefunction change in terms of the contrast of a Ramsey interferometer and control parasitic effects associated with the measurement process.
The observed back-action is in good agreement with our measurement model; this result is a necessary precursor for achieving true quantum back-action limited measurements of quantum gases.
\end{abstract}

\maketitle


Back-action limited weak measurements are essential for advancing quantum technologies, enable new probes of quantum systems, and offer new ways to understand the measurement process.
Most quantum technologies simultaneously require quantum limited measurements and feedback control to establish and maintain quantum coherence and entanglement, with applications ranging from quantum state preparation\cite{Degen2017, Polzik_RevModP2010} to quantum error correction\cite{Terhal2015}.
Even without feedback, system dynamics combined with weak measurements can lead to entangled states in the thermodynamic limit\cite{Polzik_PRL2011,GullansPRX_20,Crystal_22,Yao_PRL2022}.
Large-scale applications of these capabilities hinge on understanding system-reservoir dynamics of many-body quantum systems, whose Hilbert space grows exponentially with system size.
Ultracold atoms, a workhorse for quantum simulation\cite{DalibardBloch2008_RMP,Muller2012}, are an ideal platform for studying the system-reservoir dynamics of large-scale many-body systems.

Weakly measured quantum systems can be understood using the robust framework of quantum trajectories\cite{Carmichael1993,Molmer1993}.
In these descriptions, the system and a larger reservoir interact and become weakly entangled, at which point the reservoir is projectively measured.
This destroys the system-reservoir (SR) entanglement and leads to a change in the system's wavefunction.
We develop such a measurement model to study the interplay between the system-reservoir interaction, the scattered light, and the post-measurement system state. 

Very far from atomic resonance light Rayleigh-scatters from atomic ensembles, changing the incident light's wavevector in proportion to the Fourier transform of the atomic density distribution.
The straightforward interpretation of back-action resulting from scattered photons makes quantum trajectories an ideal tool for both intuitively and quantitatively understanding the system-reservoir interaction.
When the reservoir-measurement outcomes are rejected, quantum trajectories methods form a specific physically motivated ``unraveling'' of the master equation\cite{Molmer1993}.
In the quantum problem, light scattering gives information both about the expectation value of the density---essentially classical scattering---as well as quantum fluctuations, which contribute to spontaneous emission.
Quite recently a trio of papers observed the predicted suppression of light-scattering from deeply degenerate Fermi gases\cite{Amita2021, JunYe2021, Margalit2021} as well as amplification from ultracold Bose gases\cite{Ketterle2022_Stimulation}; these effects result from scattering atoms into occupied quantum states.

Ultracold atoms have multiple well-established ``non-destructive'' measurement techniques\cite{Andrews1996,HigbieSadler2005, Ramanathan2012,Freilich1182,Gajdacz2013,Altuntas2021}.
While backaction-induced heating of a single motional degree of freedom of a BEC was observed in a single mode optical cavity\cite{Murch2008a}, previous demonstrations of such methods with spatial resolution did not quantify quantum back-action.

\begin{figure*}[tb]
\begin{center}
\includegraphics[]{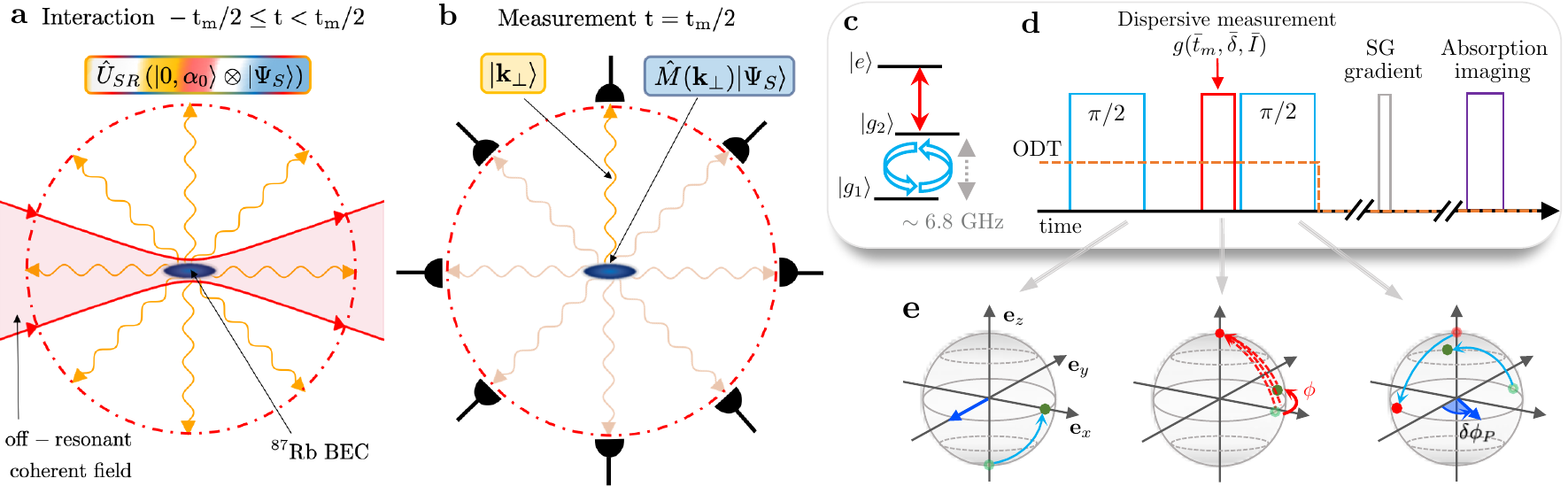}
\end{center}
\caption{{\bfseries Photodetection measurement model and Ramsey interferometry.}
{\bfseries a} Interaction. 
A Bose-Einstein condensate (BEC) illustrated in blue, is illuminated with far-detuned laser light (red) for a time $t_{\rm m}$ and scatters light (wiggly lines) into both occupied (red) and reservoir (orange) modes.
{\bfseries b} Measurement. 
The reservoir modes are projectively measured by an array of photo-detectors encompassing $4\pi$ steradians yielding the outgoing wavevector and polarization.
{\bfseries c} Level diagram.
{\bfseries d} Time sequence for Ramsey interferometry.
An initial $\pi/2$  microwave pulse (light blue) is followed by a $15\ \mu{\rm s}$ evolution period; then the $t_{\rm m} = 20~\mu{\rm s}$ off-resonant light pulse (red) is applied, and after a $5\ \mu{\rm s}$ delay (giving a total $T = 40\ \mu{\rm s}$ free evolution time) the Ramsey sequence is completed with a second $\pi/2$ pulse (light blue).
The optical dipole trap (ODT), denoted by the orange dashed line, is extinguished immediately following the Ramsey sequence.
A Stern-Gerlach (SG) gradient (grey) is applied during time-of-flight (TOF) and the final density is detected using absorption imaging (purple).
{\bfseries e}  Bloch sphere depiction of Ramsey interferometry. 
The dark blue arrows depict the axes of rotation for each microwave pulse and the light blue arrows mark the associated trajectories.
The green circles show the coherent evolution during each step of our sequence. 
Red arrows depict evolution associated with the measurement pulse with the solid curve resulting from the Stark shift and dashed curves resulting from measurement back-action.
The red circles are the states that were measured to be in $\ket{g_2}$.
Translucent (solid) symbols indicate the initial (final) state.
}
\label{Fig:MeasurementModel}
\end{figure*}

Here we characterize measurement back-action in atomic Bose-Einstein condensates (BECs), weakly interacting with a far-from resonant laser beam.
The information extracted by light-scattering can be treated as a quantum measurement process where the scattered light is detected by the environment [Fig. \ref{Fig:MeasurementModel}a-b], and we--the observer--detect only the resulting back-action on the system.
The wavefunction change is quantified by the phase shift and contrast of a Ramsey interferometer.
In our Ramsey interferometer [Fig.~\ref{Fig:MeasurementModel}c], spontaneously scattered light measures atoms to be in the detected spin state, thereby breaking its coherence and reducing the interferometer contrast.
We further distinguish between non-destructive measurements (where the system is apparently undisturbed) and back-action limited measurements (where observed quantum projection noise dominates the change in the post-measurement state).
We systematically control for two stray effects that otherwise lead to excess excitation or loss: inhomogeneities in the probe beam, and a weak optical lattice from weak back-reflections of the probe beam.
We explore a third systematic effect: light induced collisions---intrinsic atomic processes---that were found to have limited impact on our Ramsey data.
We demonstrate that these technical artifacts can be eliminated, bringing the observed back-action into agreement with our measurement model.

\section*{Results}

\subsection*{Quantum trajectories model}
\label{sec:MeasurementModel}

We consider a weakly interacting atomic BEC (the system) dispersively coupled to the optical electric field $\hat{\bf E}({\bf x},t)$ (the reservoir) by the ac Stark shift with interaction picture Hamiltonian
\begin{align}
\hat H_{\rm SR}(t) &=\! \int\! \frac{d^3 {\bf x}}{\hbar\Delta} \hat n_{\rm g}({\bf x})\otimes
[\hat{\bf E}({\bf x},t)\!\cdot\!{\bf d}_{\rm ge}][{\bf d}_{\rm ge}^{\ast}\!\cdot\!\hat{\bf E}^\dagger({\bf x},t)].
\label{Eqn:acStark_Hamiltonian}
\end{align}
Here $\hat n_{\rm g}({\bf x}) = \hat{b}_{\rm g}^{\dagger}({\bf x}) \hat{b}_{\rm g}({\bf x})$ is the atomic density operator in terms of the bosonic field operators $\hat{b}_{\rm g}({\bf x})$ for ground state atoms at position ${\bf x}$; ${\bf d}_{\rm ge}$ is the dipole matrix element for transitions between ground and excited state atoms with energy difference $\hbar \omega_{\rm ge}$; lastly, $\Delta = \omega_0 - \omega_{\rm ge}$ is the detuning from atomic resonance of a probe laser with frequency $\omega_0$.

For $|\Delta|\ll\omega_{\rm ge}$, the optical electric field operator is
\begin{align}
\hat {\bf E}({\bf x},t) &= i \sqrt{\frac{\hbar \omega_{\rm ge}}{2 \epsilon_0}}  \sum_{\sigma} \int\!\frac{d^3{\bf k}}{(2\pi)^3} \hat a_\sigma({\bf k}) \boldsymbol{\epsilon}_{\sigma}({\bf k}) e^{i ({\bf k}\cdot {\bf x}-c|{\bf k}| t )},
\label{Eqn:E_field_}
\end{align}
expressed in terms of field operators $\hat{a}_{\sigma}({\bf k})$ describing states with wavevector ${\bf k}$ and polarization $\sigma$.
Here, $c$ is the speed of light; $\epsilon_0$ is the electric constant; and $\epsilon_{\sigma}({\bf k})$ are a pair orthogonal polarization vectors transverse to ${\bf k}$, labeled by $\sigma = \pm$. 
Figure~\ref{Fig:MeasurementModel}a depicts the full system-reservoir coupling scheme with the BEC interacting with outgoing transverse modes and a probe laser in mode $({\bf k}_0,\sigma_0)$ for a duration $t_{\rm m}$.

During this time the atomic ensemble scatters monochromatic light into outgoing modes of wavevector ${\bf k}_\perp$ with coupling strength
\begin{align}
g_\sigma({\bf k}_\perp) & \equiv -i \left(\frac{\omega_{\rm ge}}{2\hbar \epsilon_0} \right)^{1/2} \left[{\bf d}_{\rm ge} \cdot \boldsymbol{\epsilon}_\sigma({\bf k}_\perp)\right].
\label{Eqn:Couple_Strength}
\end{align} 
Since each outgoing mode is in a specific polarization state ${\epsilon}({\bf k}_\perp)$ the polarization subscript is redundant.
 
Assuming that the probe laser of wavelength $\lambda$ occupies a single optical mode $({\bf k}_0, \sigma_0)$ with $\kr \equiv |{\bf k}_0| = 2\pi/\lambda$, we make the replacement $\hat a_\sigma({\bf k}) \rightarrow \delta({\bf k}-{\bf k}_0) \delta_{\sigma,\sigma_0} \alpha_0 + \hat a_\sigma({\bf k})$, which describes a coherent driving field with amplitude $\alpha_0$.
In this expression the modes $\hat a_\sigma({\bf k})$ are initially empty.
This replacement allows us to expand Eq.~\eqref{Eqn:acStark_Hamiltonian} in decreasing powers of the large parameter $\alpha_0$.
The leading term describes the ac Stark shift, and the next term 
\begin{align*}
\hat H_{\rm eff} &= \frac{\hbar P_{\rm e}^{1/2}}{(ct_{\rm m})^{1/2}}\! \oint_{k_0}\! \frac{d^2 {\bf k}_\perp}{(2\pi)^2}
g^*({\bf k}_\perp) 
\hat{n}_{\mathcal{F}}({\bf k}_\perp\!-{\bf k}_0)\hat a^\dagger({\bf k}_\perp) + \rm{H.c},
\end{align*}
describes scattering from the probe field into outgoing modes by any structure in the atomic density, with Fourier components
\begin{align*}
\hat{n}_{\mathcal{F}}({\bf k}_\perp-{\bf k}_0) = \int \frac{d^3 {\bf k}}{(2\pi)^3} \hat b^\dagger[{\bf k} - ({\bf k}_\perp-{\bf k}_0)] \hat b({\bf k}).
\end{align*}
Here $P_{\rm e} = |\alpha_0 g_{\sigma_0}({\bf k}_0)|^2 / \Delta^2 $ is the excited state occupation probability.
In the far-detuned limit, the outgoing wavenumber is fixed at $k_0$ leading to the surface integral over the sphere of radius $k_0$.

We model the larger environment as performing measurements on the outgoing light in the far-field with an ideal photo detection process, a strong measurement of the photon density $\hat{a}^\dagger({\bf k}_\perp) \hat{a}({\bf k}_\perp)$ [Fig. \ref{Fig:MeasurementModel}b]. 
In the abstract, this process begins with the combined system reservoir state $\ket{0}\otimes\ket{\Psi_{\rm S}}$, describing a reservoir with no photons but with the system in an arbitrary state.
This state evolves briefly for a time $t_{\rm m}$ via the time evolution operator $\hat{U}_{SR}(t_{\rm m}) = \mathcal{T} \exp\left[-i \int_{-t_{\rm m}/2}^{t_{\rm m}/2} \hat{H}_{\rm {eff}}(t) d t/\hbar\right]$.
This entangles the system and reservoir; as depicted in Fig.~\ref{Fig:MeasurementModel}a amplitude can be present in every reservoir mode prior to measurement by the environment.

\subsection*{Photodetection}
\label{sec:PhotoDetector_Theory}

We turn to the photodetection model shown in Fig.~\ref{Fig:MeasurementModel}b.
In this case, the measurement of the reservoir collapses the superposition by measuring either no photons or a single photon in final state $\ket{{\bf k}_\perp}$.
The back-action of this measurement is described by a conditional change in the system wavefunction $\ket{\Psi_S^\prime }= \M \ket{\Psi_S}$, an operation described by Kraus operator $\M = \bra{{\bf k}_\perp} \hat{U}_{SR}(t_{\rm m})\ket{0}$.
Taken together this schema is a generalized measurement of the system effected by projective measurements on the reservoir.

In the limit of small $t_{\rm m}$, such that at most one photon is scattered, we obtain the Kraus operator
\begin{align}
\M &= -i P_{\rm e}^{1/2}\left(\frac{t_{\rm m}}{c}\right)^{1/2} 
g^*({\bf k}_\perp)\hat{n}_{\mathcal{F}}({\bf k}_\perp-{\bf k}_0) 
\label{Eqn:Kraus_op}
\end{align} 
describing the recoil of the system from momentum-conserving scattering out of every occupied state.

The Kraus operator contains information both about the change in the system as well as the probability density
\begin{align}
{P}({\bf k}_\perp) & \equiv \bra{\Psi_S}\Mt\M\ket{\Psi_S} \label{eq:ScatProb}\\
&= \frac{t_{\rm m} P_{\rm e}}{c} |g({\bf k}_\perp)|^2
\bra{\Psi_S}\left|\hat{n}_{\mathcal{F}}({\bf k}_0 - {\bf k}_\perp) \right|^2\ket{\Psi_S} \nonumber 
\end{align}
that this change occurred.
Bringing $|\hat{n}_{\mathcal{F}}|^2$ into a normal-ordered form shows that the scattering probability has two contributions.
For a BEC with condensate mode $\tilde\psi({\bf k})$  the scattering probability is
\begin{align*}
{P}({\bf k}_\perp) &= \frac{t_{\rm m} P_{\rm e}}{c} |g({\bf k}_\perp)|^2 N \left[(N-1) \left| n_{\mathcal{F}}({\bf k}_0 - {\bf k}_\perp) \right|^2 + 1 \right],
\end{align*}
where in analogy with the operator expression, $n_{\mathcal{F}}({\bf k}_0 - {\bf k}_\perp)$
describes the Fourier components of the probability-density.
The first term describes collective scattering from the overall density profile (including thermal fluctuations), a.k.a. classical scattering\cite{BornWolf1999Book}, while the second results from scattering from quantum fluctuations, here giving rise to spontaneous emission.
For extended systems such as our BEC, the collective term is dominated by small angle forward scattering while the spontaneous term is nominally isotropic.
Notably, this result illustrates that the ratio between collective and spontaneous scattering depends on $N$ but not the measurement parameters.

Integrating over the final ${\bf k_\perp}$ states gives $P_{\rm tot} = P_{\rm col} + P_{\rm sp}$ with the spontaneous scattering probability $P_{\rm sp} = \Gamma t_{\rm m} P_{\rm e} = g^2 / 8$.
We introduced an overall measurement strength $g = \sqrt{\bar t_{\rm m} \bar I} / \bar\delta$ in terms of dimensionless: time $\bar t_{\rm m} = \Gamma t_{\rm m}$ scaled by the natural linewidth $\Gamma$; detuning $\bar\delta = \Delta / \Gamma$ in units of $\Gamma$; and laser intensity $\bar I = I / I_{\rm sat}$ in units of the saturation intensity $I_{\rm sat}$. 
Thus when $g = \sqrt{8}$ each atom will have on average spontaneously scattered a single photon (The relation between $g$ and the signal to noise ratio of a measurement outcome is briefly discussed in Supplementary Note 1.).

In experiment, a single measurement pulse can lead to thousands of photodetection events, each described by a Kraus operator.
The concatenation of many such Kraus operators---one for each scattering event---describes the evolution of our system.
By contrast with master equation methods that trace out the environment, quantum trajectories approaches predict individual measurement outcomes and the associated back-action, drawn from a suitable statistical distribution.
Thus, the final post-measurement state can be predicted given an experimentally observed measurement record.
For ensemble averaged predictions, our technique and standard methods such as those used in Appel et al.\cite{Appel2009a} give the same results.
We compare the predictions of this theoretical description with an observable, contrast in a Ramsey interferometer, that does not rely on knowledge of the specific quantum trajectory that the system followed.

\subsection*{Experimental system}\label{sec:ExpOverview}

Our experiments started with highly elongated $^{87}$Rb BECs prepared in a crossed optical dipole trap (ODT) with frequencies $(\omega_x, \omega_y, \omega_z) = 2\pi \times \left[9.61(3), 113.9(3), 163.2(3) \right]\ {\rm Hz}$ in the $\ket{g_1}\equiv\left|F = 1, m_{F} = 1\right\rangle$ electronic ground state
(All uncertainties herein reflect the uncorrelated combination of single-sigma statistical and systematic uncertainties).
This trap configuration yielded condensates with $N_{\rm c}=0.70(15)\times10^5$ atoms\cite{Dalfovo1999,Castin1996}, condensate fraction $R_{\rm c} = 78(3)\%$, and chemical potential $\mu=h\times0.76(6)\ {\rm kHz}$.
We drove transitions between $\ket{g_1}$ and $\ket{g_2}\equiv\ket{F=2,m_F=2}$ using an $\approx 6.8\ {\rm GHz}$ microwave magnetic field with Rabi frequency $\approx 7.5\ {\rm kHz}$.

In our experiments we illuminated the BEC {\it in situ} with an off-resonant probe laser beam that drove the $\ket{g_2}$ to $\ket{e}\equiv\ket{F^\prime=3,m_F^\prime=3}$ ground to excited state transition.
This probe laser was blue detuned by $0 < \bar\delta < 317$, and had intensity $\bar I \lesssim 10$.
We theoretically describe the light scattered at large angle as being subsequently projectively measured by the environment, as described above.
We then detected the post-measurement density distribution using absorption imaging after a longer $20\ {\rm ms}$ TOF during which a Stern-Gerlach gradient spatially separated the $\ket{g_1}$ and $\ket{g_2}$ components. 

\subsection*{Detecting Wavefunction Change via Ramsey Interferometry}
\label{sec:RamseyLightShift}

\begin{figure}[tb]
\includegraphics{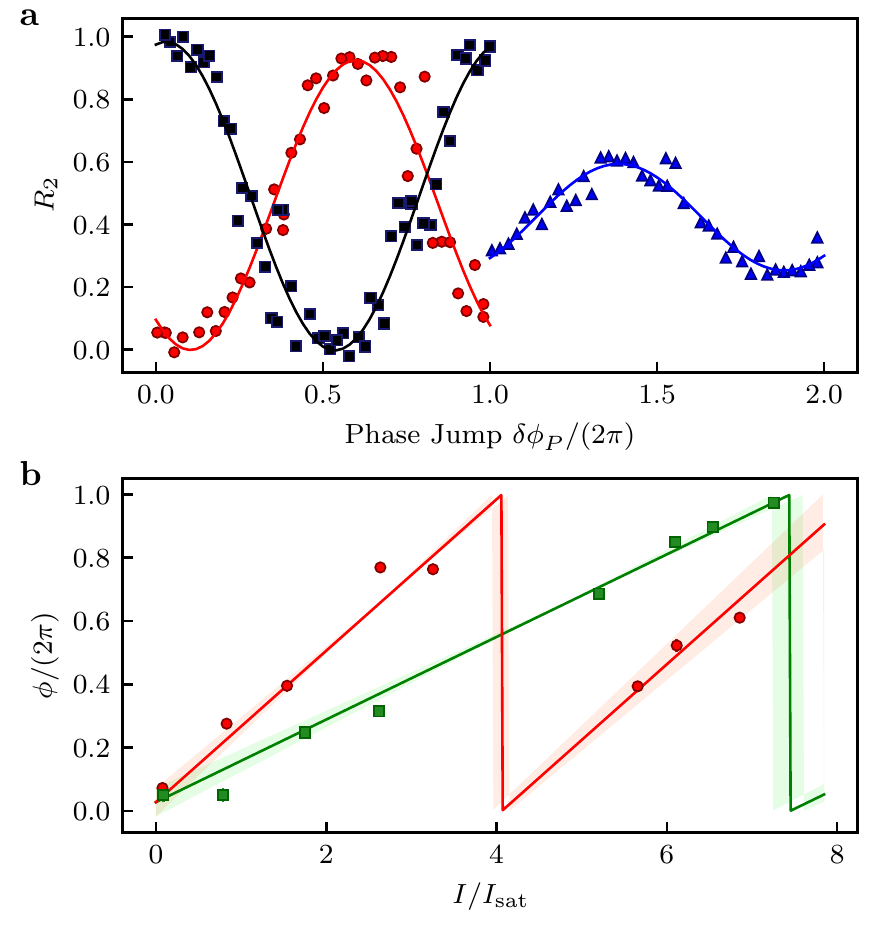}
\caption{{\bfseries Ramsey interferometry data.} 
{\bfseries a} Ramsey oscillation without (black squares) and with the light pulse at $\bar\delta = 63.4$ and $\bar I \approx 2 $ (red circles) and $\bar I \approx 7 $ (blue triangles), taken {\it in situ} as the phase jump $\delta \phi_{\rm P}$ between the two pulses is varied.
Solid curves are fits to the equation given in the text.
Blue Ramsey data $\delta\phi_{P}$ values are deliberately shifted by $2\pi$ to visualize the reduction in contrast and the larger phase shift. 
{\bfseries b}  Optically induced phase shift $\phi$ as a function $\bar I$ at $\bar\delta = 63.4$ (red circles) and $\bar\delta = 116.2$ (green squares).
The same-color lines are fits to $-V_{\rm ac} t_{\rm m} / \hbar\ {\rm mod}\ 2\pi$.
Shaded regions indicate the $\pm1\sigma$ statistical uncertainty range.
}
\label{Fig3:LightShift_Bloch}
\end{figure}

We characterize the light matter interaction, as well as back-action, predicted by our quantum trajectories model using Ramsey interferometry (RI).
Our Ramsey interferometer [Fig.~\ref{Fig:MeasurementModel}d-e] commenced with a resonant microwave pulse driving a $\pi/2$ rotation about $\ey$, taking the atoms from $-\ez$ (in $\ket{g_1}$) to $\ex$.
Then during the free evolution time we applied the probe laser detuned by $\bar\delta$ from the $\ket{g_2}$ to $\ket{e}$ transition for a time $t_{\rm m}$; the resulting ac Stark shift drove a rotation about $\ez$ by $\phi$ (solid red arc).
A second microwave pulse drove a $\pi/2$ rotation about an axis rotated by $\delta\phi_{\rm P}$ at which time we measured the final populations $N_1$ and $N_2$ in $\ket{g_1}$ and $\ket{g_2}$ respectively in TOF, giving the fraction in $\ket{g_2}$ as $R_2 = N_2/ (N_1 + N_2)$.
The black data (squares) in Fig.~\ref{Fig3:LightShift_Bloch}a, taken with the probe laser off, shows that the resulting fractional population $R_2$ is cosinusoidal, and the red data (circles), with the probe on, is phase shifted (from the ac Stark shift on $\ket{g_2}$).
We obtain the phase shift $\phi$, contrast $A$, and center shift $b$ with fits to $R_2 = [1 + A\cos(\delta\phi_P+\phi)]/2+b$.

The RI phase shift is a direct measure of the differential phase acquired during free evolution, here $-V_{\rm ac} t_{\rm m} / \hbar$ from the ac Stark shift of $\ket{g_2}$ due to the probe beam, with $V_{\rm ac} = \Gamma \bar I / (8 \bar\delta)$. 
The Stark shift of $\ket{g_1}$ is a small contribution that we do not include in our fits.
The data in Fig.~\ref{Fig3:LightShift_Bloch}b was taken at $\bar\delta = 63.4$ and $116.2$ (circles and squares respectively).
As expected the slope is larger for smaller $\bar\delta$, but in both cases the acquired phase can exceed $2\pi$ at which point it wraps back to zero.
The intensity of the probe laser is difficult to obtain in-vacuo\cite{Reinaudi2007,Hueck2017}; however, fitting $t_{\rm m} V_{\rm ac}$ to these data gives a direct calibration of the laser intensity, providing a conversion between our camera signal and $I_{\rm sat}$ with $ < 5\ \%$ fractional uncertainty.
We imaged the {\it{in situ}} probe beam (with no atoms present) on a charge coupled device camera to obtain the local probe intensity (in arbitrary camera units) at the location of the BEC. 
Further details are described in 
Altuntas et al.~\cite{Altuntas2023_Isat}.
The solid lines in Fig.~\ref{Fig3:LightShift_Bloch}b are the result of this fitting process.

Figure~\ref{Fig3:LightShift_Bloch}a shows a second effect of increasing measurement strength (blue data): the Ramsey contrast decreases with increasing measurement strength, implying that the post-measurement many-body wave function is not described by a coherent superposition of $\ket{g_1}$ and $\ket{g_2}$.

Our measurement model predicts this effect: as illustrated in the middle Bloch sphere in Fig.~\ref{Fig:MeasurementModel}e, each time a photon is spontaneously scattered and detected by the environment, the wavefunction of a single atom collapses into $\ket{g_2}$ (along $\ez$), losing any coherence with $\ket{g_1}$ (red dashed arrows).
The second $\pi/2$ pulse always returns that atom to the equator of the Bloch sphere, reducing the contrast by $1/N$.
In this situation, the per-atom probability of scattering a single photon at large angle is $g^2 / 16$ (see Supplementary Note 2 for the complete calculation).
By contrast for collective scattering (generally at small-angle), a detected photon scattered off of the global density distribution yields Mössbauer-like collective back-action and no reduction in contrast.
As a result, the change in contrast measures the number of spontaneously scattered photons\cite{Appel2009a}.

The ideal Ramsey interferometry scheme presented in Fig.~\ref{Fig:MeasurementModel}d is sensitive to additional systematic effects leading to contrast reduction.
In the following sections we identify such factors, and develop an enhanced RI scheme that detects the post-measurement wavefunction change in agreement with the theoretical prediction.

\begin{figure}[htb]
\begin{center}
\includegraphics[]{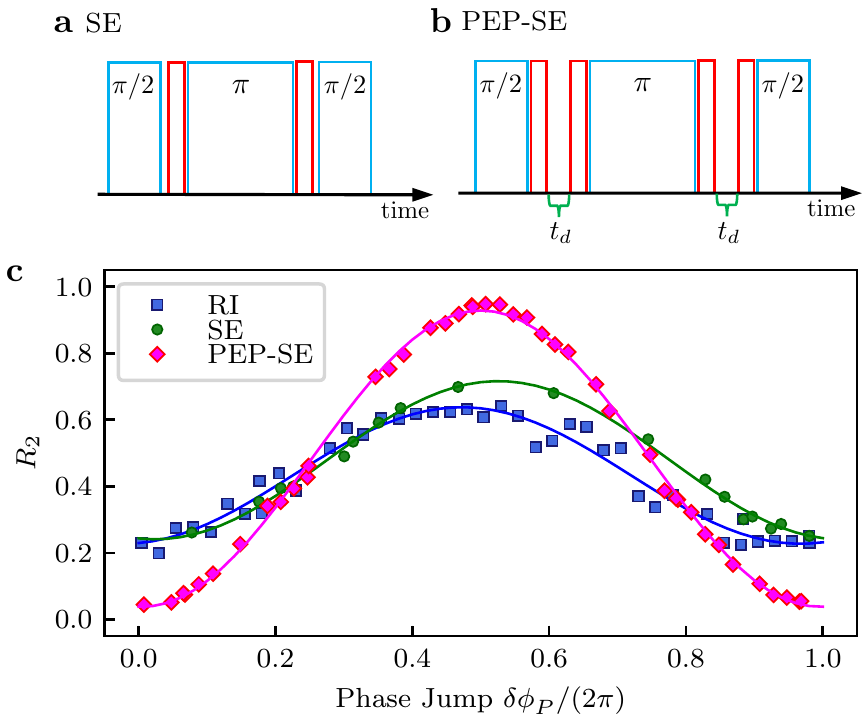}
\end{center}
\caption{{\bfseries Improved Ramsey interferometers.} 
{\bfseries a}, {\bfseries b} Pulse sequences for spin-echo (SE) and pulse-evolve-pulse with spin-echo (PEP-SE) Ramsey interferometers. 
The red blocks denote the dispersive measurement pulses, and blue bars indicate the microwave pulses. 
{\bfseries c} Interferometer signal measured using: a conventional Ramsey sequence (blue squares), a spin-echo Ramsey sequence (green circles), and a pulse-evolve-pulse with spin-echo sequence (magenta diamonds). 
All measurements were performed {\it in situ} at $\bar\delta = 63.4$ with $\bar I \approx 7 $ yielding $g \approx 1$. 
}
\label{Fig3:RI_PulseSch_PEPSE}
\end{figure}

\subsection*{Spin-echo Ramsey Interferometer} \label{Section:SpinEchoTOF}

Spatial inhomogeneities in the probe beam as well as near-dc magnetic field noise can reduce the RI contrast.
In the first case, the resulting position-dependent ac Stark shift imprints spatial structure to the RI phase $\phi$, thereby reducing the spatially averaged contrast. 
Second, because the $\ket{g_1}$-$\ket{g_2}$ transition is first-order sensitive to the external magnetic field, the RI contrast is reduced when field noise randomly shifts the resonance condition between different repetitions of the experiment.

We added a spin-echo pulse to our interferometer [see Fig.~\ref{Fig3:RI_PulseSch_PEPSE}a]
to compensate for both of these parasitic effects.
As Fig.~\ref{Fig3:RI_PulseSch_PEPSE}c shows, the noise in the spin-echo signal (circles) is reduced compared to the standard RI measurement (squares).
Although the measurement noise is reduced, the contrast with spin echo is unchanged (Fig.~\ref{Fig4:RI_Contrast_PulseSch}a-b), leaving the substantial disagreement with our theory prediction (black curve) due to the systematic factor we report next.

\subsection*{Ramsey Interferometer with Split Measurement Pulses} \label{Section:Lattice}

\begin{figure*}[tb]
\begin{center}
\includegraphics[]{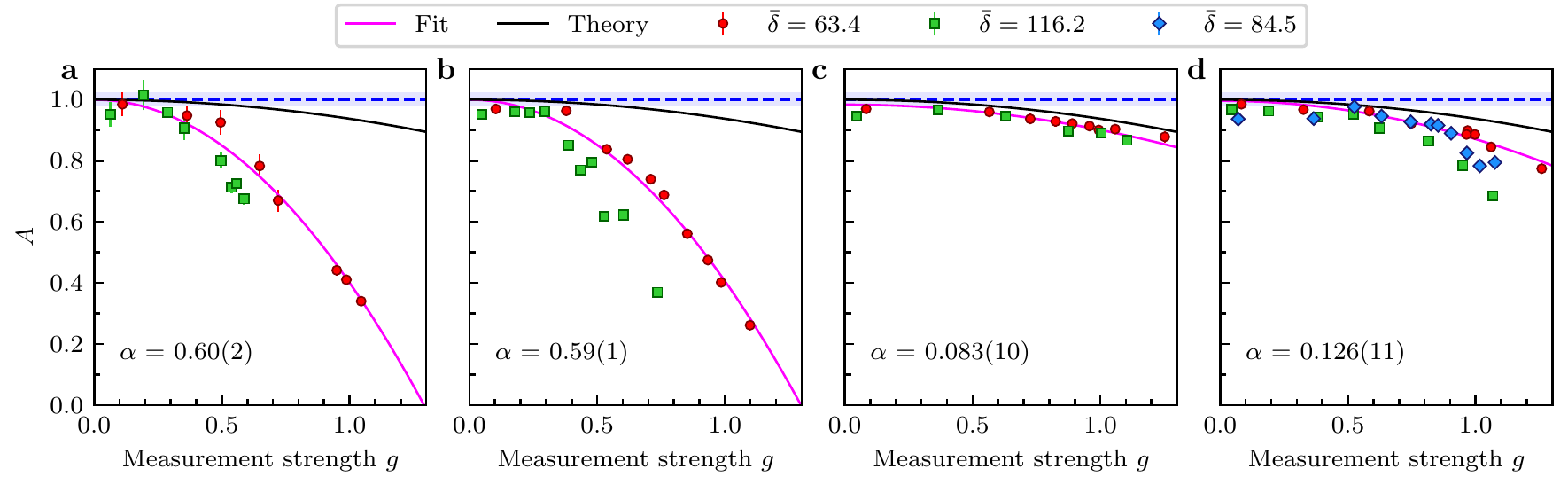}
\end{center}
\caption{
{\bfseries Ramsey interferometer contrast $A$ dependence on the measurement strength $g$ for different schemes.}
{\bfseries a} {\it In situ} Ramsey interferometry measurements.
{\bfseries b} {\it In situ} Ramsey interferometry with spin-echo measurements.
{\bfseries c} {\it In situ} Pulse-evolve-pulse with spin-echo Ramsey interferometry measurements.
{\bfseries d} Pulse-evolve-pulse with spin-echo sequence after 2 ms TOF measurements.
The horizontal blue lines show the RI contrast observed without the measurement pulse with the shaded regions indicating the $\pm1\sigma$ statistical uncertainty range.
The black curves plot the prediction of our photodetection model.
The pink curves depict a fit of the $\bar\delta = 63.4$ data to $A = A_0-\alpha g^2$ with each best-fit alpha value quoted on the respective figure.
}
\label{Fig4:RI_Contrast_PulseSch}
\end{figure*}

Contrary to our predictions, Fig.~\ref{Fig4:RI_Contrast_PulseSch}a-b show that the contrast depends on probe detuning (green squares versus red circles).
This difference signifies the presence of the second parasitic effect: a weak optical lattice generated by the probe beam interfering with its retro-reflections off subsequent optical elements.
The probe beam is nearly perfectly concentric with our imaging system and intersects each optical element at normal incidence. 
While it is common practice in optical setups to slightly tilt optical elements to eliminate back-reflections, in the high-resolution imaging context optimized alignment is a necessary condition for minimizing optical aberrations.

As a result, each probe pulse corresponds to the sudden application of a lattice potential.
Weak lattices create populations in matterwave diffraction orders with momentum $\pm 2 \hbar \kr$.
In principle a suitable spin-echo sequence could remedy this, nonetheless, the rapidly moving diffracted atoms experience different lattice potentials during our first and second pulses precluding effective cancellation.

Instead we extended the ideas in Wu et al.\cite{Wu2005} and Herold et al.\cite{Trey_2012PRL} by splitting each probe pulse into two pulses of duration $t_{\rm p} = 8.2~\mu{\rm s}$ spaced in time by a carefully chosen $t_{\rm d} = 25.6~\mu{\rm s}$ of free evolution, essentially unwinding the phase imprinted by the lattice (see Supplementary Note 3).
Fig.~\ref{Fig3:RI_PulseSch_PEPSE}b shows such a pulse-evolve-pulse with spin-echo (PEP-SE) sequence.
The near-full contrast magenta Ramsey fringe in Fig.~\ref{Fig3:RI_PulseSch_PEPSE}c results from this PEP-SE sequence applied {\it in-situ} for $g\approx 1$.
As seen in Fig.~\ref{Fig4:RI_Contrast_PulseSch}c, there is negligible difference in the extracted contrast between measurements at the same $g$ value but with different probe detunings (squares and circles) further confirming control over systematic effects. 
The PEP-SE Ramsey contrast is in good agreement with our theoretical model (black curve) and provides a mechanism for identifying the regime of back-action limited measurements of ultracold gases.
In order to obtain a quantitative metric for comparison with theoretical prediction, we fit the $\bar\delta = 63.4$ data in $g\leq 1$ to $A = A_0-\alpha g^2$, where $A_0$ describes a small overall reduction in contrast. 
PEP-SE scheme measurements yield $\alpha = 0.083(10)$, which is in good agreement with the theoretical prediction $\alpha_{\rm{th}} = 1/16 \approx 0.063$.

\subsection*{Light induced collisions} \label{Section:PA_losses}

\begin{figure}[hbt]
\begin{center}
\includegraphics[]{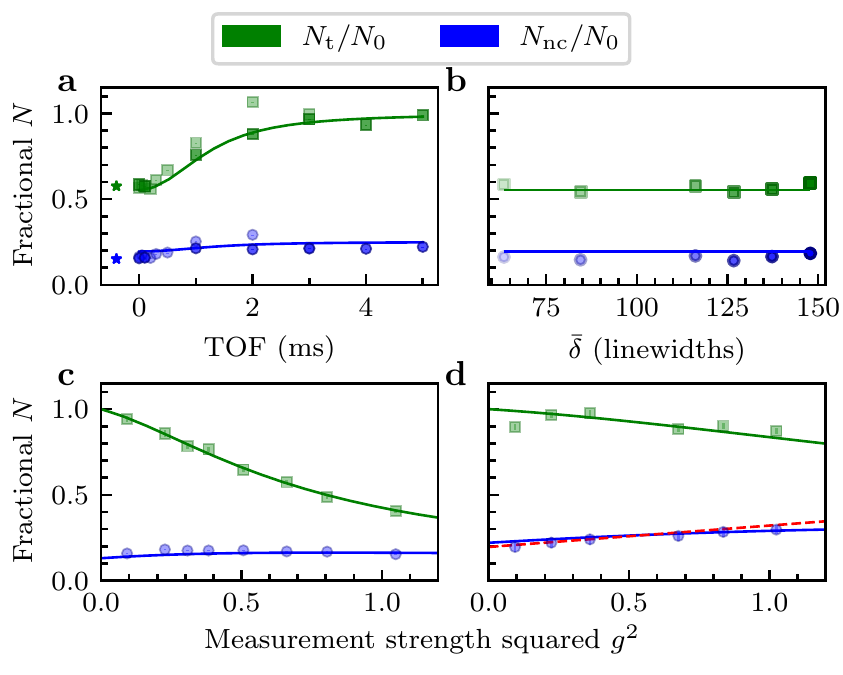}  
\end{center}
\caption{{\bfseries Light induced collisions.}
Fractional total number of atoms (green squares) and number of atoms outside the BEC but within a 1 recoil momentum circle (blue circles) following measurement.
The symbols mark experimental data while the curves are the result of our 2-body model.
In all cases, the opacity of the points reflects the detuning.
{\bfseries a} Dispersive measurements at two different probe detunings ($\bar\delta = 84.5$ and $\bar\delta = 126.7$) after a variable short time-of-flight (TOF).
All measurements were at measurement strength $g \approx 1$, which was attained by adjusting intensity to $\bar{I}\approx 9.5$ and $\bar{I}\approx 21$ respectively.
Star symbols mark {\it in situ} measurements with the optical dipole trap on (plotted at negative TOF for display purposes). 
{\bfseries b} {\it In situ} measurements at different $\bar\delta$ all with $g \approx 1$. 
Measurement time was $t_{\rm m} = 25\ \mu{\rm s}$ in (a) and (b).
{\bfseries c} and {\bfseries d} Loss as a function of $g^2$ for measurements made {\it in situ} in (c) and with a $2\ {\rm ms}$ TOF in (d).
The red dashed line in (d) plots the expected $g^2/8$ light-scattering behavior.
Both cases were at $\bar\delta = 84.5$ and $\bar{I}\approx 9.5$ with $t_{\rm m}$ varied from $4\ \mu{\rm s}$ to $36\ \mu{\rm s}$. }
\label{Fig2:Losses}
\end{figure}

We used the post-measurement atom number as an auxiliary probe of measurement back-action and found that, although photoassociation (PA) is suppressed at blue detuning, at our high {\it in situ} atomic densities of $1\times10^{14}\ {\rm cm}^{-3}$, light induced collisions lead to rampant atom loss\cite{Ketterle2022_Stimulation}.
We quantify the importance of these losses by preparing BECs with $N_0$ total atoms in $\ket{g_2}$ and measuring fractional change in total atom number $N_{\rm t}/N_0$ and in uncondensed number $N_{\rm nc}/N_0$.
$N_{\rm nc}/N_0$ counts both thermal atoms as well as atoms that have undergone large-angle light scattering.

Fig.~\ref{Fig2:Losses}a confirms that this is a 2-body process by reducing the atomic density with a short TOF.
We find that the losses rapidly drop starting at $t_{\rm TOF}\approx0.5\ {\rm ms}$ (when mean-field driven expansion becomes significant) and vanish after $3\ {\rm ms}$ (at which time the density has dropped by a factor of nearly 20).
We also investigated another potential loss mechanism due to two-color PA resulting from the combination of the intense dipole trapping beam and the probe beam. 
Data taken just before (star symbols at negative time for clarity) and just after the ODT turn-off have no difference in loss, confirming the absence of any two-color PA effects.

Panel b, taken {\it in situ}, shows that the fractional number is independent of $\bar \delta$.  
These data were taken at constant $P_{\rm e}$ (achieved by tuning $\bar I$) and demonstrate that there are no PA resonances.
Figure~\ref{Fig2:Losses}c shows that {\it in situ} the total number drops rapidly with increasing $g^2$ while the number outside the BEC remains constant.
This verifies that the high-density BEC experiences light induced collisions while the low density thermal cloud is left mostly unchanged.
Lastly Fig.~\ref{Fig2:Losses}d plots these quantities following a $2\ {\rm ms}$ TOF, confirming the same reduced losses found in Fig.~\ref{Fig2:Losses}a.
Furthermore $N_{\rm nc}$ increases linearly with slope $g^2/8$ (red dashed curve) as expected from photon scattering. 

All of these data are well described by a 2-body loss model (solid curves), however, these simulations require a 2-body coefficient that is about $20\times$ in excess of the PA loss coefficient found in Fuhrmanek et al.~\cite{Fuhrmanek2012}.
In fact these observations reflect different processes: in the blue-detuned case light induced collision leads to rapidly accelerated atom pairs rather than PA\cite{Fung2014}.

Lastly, we note that light traversing the BEC acquires a phase shift causing the atomic cloud to act as a lens. 
When the phase shift is in excess of about $1$ radian the scattering is no longer described by our model and atomic cloud experiences excess compression, potentially enhancing 3-body loss.
The absence of $\bar\delta$ dependence in Fig.~\ref{Fig2:Losses}b affirms that effects such as this arising from the ac Stark shift do not contribute to loss.


As light assisted collisions precipitate atom loss, we added a short TOF to the spin-echo pulse-evolve-pulse Ramsey sequence to study the impact of light assisted collisions on RI contrast.
As shown in Fig.~\ref{Fig4:RI_Contrast_PulseSch}d, the contrast is modestly reduced, and as with Fig.~\ref{Fig4:RI_Contrast_PulseSch}a-b, data taken at larger detuning are impacted more significantly. 
We attribute this reduction to the changing optical intensity profile that the falling BEC experiences as it traverses different regions of the probe beam during the pulse sequence; this compromises the PEP-SE sequence.

\section*{Discussion}
\label{sec:Conc}

Even though RI contrast is a direct measure of the overall wavefunction change, our light assisted collision data show that RI contrast alone is insufficient to identify back-action dominated measurement regimes.
For our {\it in situ} results---with rampant light induced losses---photon scattering from the measurement process does not fully explain the change of the system's state. 
Consequently such measurements are not back-action limited, even in principle.
An interesting question that we did not touch on, is how light induced collisions are able to remove atoms while leaving the Ramsey contrast largely unchanged.

For the modest range of detuning explored here, the two-body loss rate scales as the excited state probability $P_{\rm e}\propto g^2$; this implies that for a target measurement strength, light induced collisions are not reduced until vastly larger detuning when this scaling breaks down\cite{Kampel2012}.
In our experiment, data taken with $g \lesssim 0.3$ (with per-atom spontaneous scattering probability $P_{\rm sp}\lesssim0.01$) had no discernible loss in Ramsey contrast or reduction in atom number: functionally non-destructive\cite{Andrews1996}.
However, our results demonstrate that such functionally non-destructive measurements can be far from quantum back-action limited.
As a consequence, back-action limited measurements of BECs can be achieved either by managing the atom density, or by careful control of molecular resonances\cite{Urvoy2019}.
In degenerate Fermi gases the Pauli pressure leads to much lower densities\cite{Ketterle2008}, typically diluted by an order of magnitude or more compared to BECs, making two-body losses less significant.

Employing the strategies identified here is necessary to achieve back-action limited measurements, and as a next step the scattered light must actually be detected.
There are multiple imaging techniques for quantum gases based on the dispersive light-matter interaction\cite{Andrews1996, Ketterle99, Anderson2001, Gajdacz2013} that in principle can give back-action limited measurement outcomes.
Implementing these requires an imaging system with minimal losses and large numerical aperture in conjunction with a high efficiency detector, as any scattered light that is not detected is effectively measured by the environment and its information lost.
Furthermore, the captured signal must lead to a faithful representation of the atomic ensemble, necessitating an imaging system with minimal or well-calibrated aberrations as we demonstrated previously\cite{Altuntas2021}.
Lastly, the initial optical field must be well known, for which techniques such as outlined here and described in more detail in Altuntas et al.~\cite{Altuntas2023_Isat}, are essential.
These physical considerations do not touch on technical matters such as calibrating the response and hardware specific noise properties of the physical detector, i.e., a charge coupled device (CCD) or complementary metal oxide semiconductor (CMOS) camera. 
Future work needs to account for these sources of technical noise. 

Looking forward, back-action limited weak measurements coupled with real-time control are enabling tools for quantum technology.
Feedback cooling is one application of closed loop quantum control, and the interplay between measurement back-action and the actual information extracted from the system limits the achievable temperature\cite{Ivanova2005,Rempe2010,Mitchell_2013}.
In addition to simply cooling into established quantum states (both weakly and strongly correlated), closed-loop feedback enables the engineering of artificial, non-local, and non-Markovian, reservoirs.
Existing proposals with engineered reservoirs show that suitable quantum jumps lead to equilibration into strongly correlated states\cite{Diehl2010a}; and schemes using feedback can generate new Mott insulating phases\cite{Young2021a} and squeezed states\cite{Walker2020a,Molmer2015}.
In the latter case a single measurement locally creates conditional squeezing that requires a second spatially resolved control pulse---conditioned on the measurement outcome---to obtain useful unconditional squeezing.
Metrological implementations would also require atoms individually confined in the sites of an optical lattice to prevent spatial diffusion and clock shifts.

In addition, weak measurements offer new ways to explore fundamental concepts in quantum mechanics.
For example, a weak measurement of strength $g$ can be decomposed into a series of $N$ sub-measurements\cite{Caves1987, Brun2002} each with strength $g/\sqrt{N}$.
In this configuration, the total outcome of these measurements recovers an individual measurement of strength $g$, but the quantum back-action of earlier sub-measurements correlates with the outcome of later sub-measurements, giving information that is erased in a single stronger measurement.
For example, correlating the outcome of two sub-measurements can isolate the measurement back-action of the first measurement.

\section*{Methods}

\subsection{Magnetic field lock}\label{app:uwaveLock}

Our interferometry measurements operate on the magnetic field sensitive $\ket{F=1,m_F=1}$ to $\ket{F=2,m_F=2}$ transition, and as a result are negatively impacted by magnetic field noise. 
To minimize any effect on contrast, we monitored the field shifts using a microwave based monitoring scheme first implemented in LeBlanc et al.\cite{LeBlanc_2013}.

Our two level system is well described by the Hamiltonian
\begin{align}
\hat H_{\rm \mu} &= \frac{\hbar}{2}
\left(\begin{array}{cc}
\Delta_\mu + \delta_\mu & \Omega_\mu \\
\Omega_\mu & -(\Delta_\mu + \delta_\mu)
\end{array}\right)\nonumber,
\end{align}
where $\Delta_\mu$ describes an unknown detuning from resonance, $\delta_\mu$ is an adjustable detuning, and $\Omega_\mu$ is the microwave Rabi frequency.

Our protocol began with optically trapped atoms just above $T_{\rm c}$ in the $\ket{F=1,m_F=1}$ hyperfine state.
We applied a microwave pulse of duration $t_{\mu}=100\ \mu{\rm s}$ and Rabi frequency $\Omega_\mu/(2\pi) \approx 0.1 /t_\mu$ detuned by $\delta_\mu/(2\pi) = 1/(2t_{\mu}) = 5\ {\rm kHz}$ from resonance and absorption-imaged the atoms transferred to $\ket{F=2,m_F=2}$ in-situ ($\approx 10\%$ fractional transfer) leaving $\ket{F=1, m_F=1}$ state atoms undisturbed.
We used these data to obtain the transferred atom number $N_+$.
Then after a $\approx 34 \ {\rm ms}$ delay, we repeated the processes with $\delta \rightarrow -\delta$, giving $N_-$. 
The delay between the transfer pulses was selected to be an integer multiple of the $T_{\rm line} = (60\ {\rm Hz})^{-1} \approx 17\ {\rm ms}$ line period.

The fractional imbalance between the transferred numbers
\begin{align}
\varepsilon &= \frac{N_+ - N_-}{N_+ + N_-} \approx -4 t_\mu\frac{\Delta_\mu}{2\pi},
\label{Eqn:ImblanaceUwave}
\end{align}
provides an error signal that can be related to any overall shift in detuning $\Delta_\mu$ (see Supplementary Figure 2 in Supplementary Note 4). 
For example $\epsilon = 0.5$ corresponds to a detuning of just $\Delta_\mu/ 2\pi \approx 1.25\ {\rm kHz}$.

We employed a two step procedure to minimize the impact of field noise during interferometry experiments.
First, prior to any measurement sequence we optimized the bias field to minimize $\delta B$.
Second, we post-selected data to exclude cases with $|\epsilon| > 0.5$; this value was determined empirically to retain most of the data while notably removing outliers in measured contrast.

\subsection{Lattice pulse sequence}
\label{app:Lattice}

An intuitive picture of our scheme for mitigating the effect of the optical lattice begins with a three-state truncation\cite{Wu2005,Trey_2012PRL} of the full lattice Hamiltonian
\begin{align}
\frac{\hat H(k)}{\Er} &= 
\left(\begin{array}{ccc}
(k+2\kr)^2 & s/4 & 0 \\
s/4 & k^2 & s/4 \\
0 & s/4 & (k-2\kr)^2
\end{array}\right),
\end{align}
describing a lattice of depth $s\Er$, with single photon recoil momentum $\hbar\kr = 2\pi\hbar/\lambda$, energy $\Er=\hbar^2 \kr^2 / (2 m)$, and time $T_0 = 2\pi\hbar/\Er\approx 265\ \mu{\rm s}$.
For atoms initially at rest, i.e. $k=0$, this is a resonant lambda coupling scheme with bright state subspace spanned by $\ket{b_0}=\ket{k=0}$ and $\ket{b_1}=(\ket{k=+2\kr}+\ket{k=-2\kr}) / \sqrt{2}$ and an uncoupled dark state $\ket{d}=(\ket{k=+2\kr}-\ket{k=-2\kr}) / \sqrt{2}$.

Since our initial state $\ket{k=0}$ is in the bright state manifold, we focus on the bright state Hamiltonian
\begin{align}
\frac{\hat H_{\rm b}(0)}{\Er} &= 
\left(\begin{array}{cc}
0 & s/(2\sqrt{2}) \\
s/(2\sqrt{2}) & 4
\end{array}\right)\nonumber \\
&= 2 \hat I + \frac{1}{2}\left[4\hat \sigma_z + \frac{s}{\sqrt{2}} \hat \sigma_x\right].
\end{align}
When the lattice is off, this Hamiltonian describes Larmour precession around $\ez$ with Rabi frequency $4\Er/\hbar$ and when the lattice is on it describes precession about $4 \ez + [s / \sqrt{2}]\ex$ with Rabi frequency $\sqrt{16 + s^2/2} \Er/\hbar$.
In the limit $s\ll4\sqrt{2}$, the axis of rotation is tipped by $\theta = 4s/\sqrt{2}$, and the Rabi frequency is nearly unchanged from $4\Er/\hbar$.
In Supplementary Note 3 Supplementary Figure 1a plots the top of the Bloch sphere with two example orbits in this limit (dashed lines), both for zero (red) and non-zero $s$ (blue).

The solid curves in Supplementary Figure 1a show the trajectory for a two pulse sequence that also returns to the origin.
In the small $s$ limit, the condition to return to the initial state is $t_{\rm d}/T_0 = 1/8 - t_{\rm p}/T_0$, where $t_{\rm d}$ is the delay time between pulses and $t_{\rm p}$ is the pulse duration.
Supplementary Figure 1b plots the probability that the final state returns to $k=0$ for a shallow lattice with $s=1$ (computed using $7$ momentum states).
The red line indicates the predicted minimum which is in good agreement with the numerically evaluated optimum configuration.

Supplementary Figure 1c plots the same quantity, now with $s=10$, showing the narrow range of parameters for which our scheme is expected to be successful.
For most parameters, the large $s$ simulation is qualitatively different from the small $s$ results, with the exception of very short pulse times and the region following our scheme.
In practice we selected $t_{\rm d} = T_0/10 = 26.5\ \mu{\rm s}$ and $t_{\rm p} = T_0/32 = 8.2\ \mu{\rm s}$, marked by the red star in Supplementary Figure 1c.

\subsection{Conventional parameters}
\label{app:g_definition}

Here we outline the relationships between conventional experimental parameters and the relatively abstract quantities employed in deriving the coupling strength $g_\sigma({\bf k}_\perp)$ in Eq.~\eqref{Eqn:Couple_Strength}.

We start with the coherent state amplitude $\alpha_0$ and relate it to the optical intensity
\begin{align}
I = \frac{1}{2}\epsilon_0 c |E|^2 = \hbar\omega_{\rm ge} c |\alpha_0|^2.
\label{Eqn:I}
\end{align}
In the second statement we inserted the expression
\begin{align}
|E|^2 = \frac{2\hbar \omega_{\rm ge} |\alpha_0|^2}{\epsilon_0}
\end{align} 
for the magnitude of the electric field.
The saturation intensity is a key metric of the light-matter interaction; for arbitrary light polarization 
\begin{align}
I_{\rm sat} &= \frac{\epsilon_0 c \Gamma^2\hbar^2}{4 |\boldsymbol{\epsilon}_{\sigma_0}({\bf k}_0)\cdot {\bf d}_{\rm ge}|^2}.
\label{Eqn:Isat_to_g}
\end{align}
As detailed in the main text, ${\bf d}_{\rm ge}$ is the dipole matrix element for transitions between the ground and excited state with energy difference $\hbar \omega_{\rm ge}$ and where ${\bm \epsilon}_{\sigma}({\bf k}_0)$ are the polarization vectors as pairs of orthogonal vectors transverse to ${\bf k}_0$. 
In terms of these parameters the transition linewidth is
\begin{align}
\Gamma &= \left( \frac{|{\bf k}_0|^3}{3\pi\hbar \epsilon_0}\right) |{\bf d}_{\rm ge}|^2.
\end{align}
We recall the standard definition for saturation intensity 
\begin{align}
\frac{I}{I_{\rm sat}} = 2\left|\frac{ \Omega}{ \Gamma}\right|^2,
\label{Eqn:Isat_standard}
\end{align}
acquired from a more traditional treatment, where
\begin{align}
\Omega = |\boldsymbol{\epsilon}_{\sigma_0}({\bf k}_0)\cdot {\bf d}_{\rm ge}| \frac{E_0}{\hbar}
\end{align}
is the Rabi frequency.
We next express $\Omega$ in terms of the coupling strength and the optical field amplitude $\alpha_0$ giving
\begin{align}
\Omega = 2 |g_{\sigma_0}({\bf k}_0) \alpha_0|. \label{Eqn:VIII}
\end{align}
These relations allow us to bridge between conventional laboratory parameters and those employed in our model. 
For example, we combine Eq.~\eqref{Eqn:I} and Eq.~\eqref{Eqn:Isat_to_g}, to obtain
\begin{align}
\bar I = \frac{I}{I_{\rm sat}}  = \frac{8 |\alpha_{\sigma_0} g_{\sigma_0}({\bf k}_0)|^2}{\Gamma^2} 
\label{Eqn:Isat_theory}
\end{align}
in agreement with Eq.~\eqref{Eqn:Isat_standard} and Eq.~\eqref{Eqn:VIII}.

We now turn to the scattering probability $P_{\rm sp} = \Gamma t_{\rm m} P_{\rm e}$, which contains the excited state probability
\begin{align}
P_{\rm e} &= \frac{|\alpha_0 g_{\sigma}({\bf k}_0)|^2}{\Delta^2}. 
\end{align}
This, along with Eq.~\eqref{Eqn:Isat_theory}, allows us to rewrite the scattering probability as
\begin{align}
P_{\rm sp} = |\alpha_0 g_{\sigma_0}({\bf k}_0)|^2 \frac{\Gamma t_{\rm m}}{\Delta^2} = \frac{\Gamma t_{\rm m}}{8}\frac{I}{I_{\rm sat}}\frac{\Gamma^2}{\Delta^2}.
\label{Eqn:Ptot_theory}
\end{align}
This expression is organized into the physically relevant dimensionless quantities $\bar t_{\rm m}$, $\bar I$ and $\bar \delta$ introduced in the main text.
In the main text, we defined the overall measurement strength using these parameters and made the choice to not include the factor of 8 so
\begin{align}
g^2 = \frac{I}{I_{\rm sat}}\frac{\Gamma t_{\rm m}}{(\Delta/\Gamma)^2} = 8 P_{\rm sp}.
\end{align}
As such, a measurement strength of $g^2 = 1$ signifies a probability of $1/8$ for an atom to scatter a single photon at large angle.

\begin{acknowledgments}
	The authors thank J. V. Porto and M. Gullans for carefully reading the manuscript.
	This work was partially supported by the National Institute of Standards and Technology, and the National Science Foundation through the Physics Frontier Center at the Joint Quantum Institute (PHY-1430094) and the Quantum Leap Challenge Institute for Robust Quantum Simulation (OMA-2120757).
\end{acknowledgments}

\bibliography{main}

\pagebreak
\widetext
\begin{center}
	\textbf{\large Supplementary Information \\
		Quantum Back-action Limits in Dispersively Measured Bose-Einstein Condensates}
\end{center}
\setcounter{equation}{0}
\setcounter{figure}{0}
\setcounter{table}{0}
\setcounter{page}{1}
\makeatletter

\renewcommand{\thepage}{S\arabic{page}} 
\renewcommand{\figurename}{}
\renewcommand{\thefigure}{{\bf{Supplementary Figure \arabic{figure}}}}%
\renewcommand{\thesection}{Supplementary Note \arabic{section}}%
\renewcommand{\theequation}{S\arabic{equation}}
\renewcommand{\refname}{Supplementary References}
\renewcommand{\citenumfont}[1]{S#1}

\section{M\lowercase{easurement strength and signal to noise}}\label{supp:SNR}

In this section we briefly consider the relation of the measurement strength $g$ to signal to noise ratio (SNR) for the determination of atom number using either forward-directed collectively scattered light or spontaneously scattered light.
For standard imaging, directly detecting the collectively scattered light would typically be performed using dark-field imaging (which in principle has the same SNR as phase-contrast imaging), and detecting the spontaneously scattered light would be realized via fluorescence imaging.
In principle there is no hard distinction between these techniques since large numerical aperture imaging systems can capture significant contributions from both.
For practical experiments with degenerate gases the forward scattered light dominates.

For a Bose-Einstein condensate (BEC), and for simplicity assuming an isotropic (rather than dipole) scattering distribution, the scattering probabilities are
\begin{align}
	P_{\rm tot} = P_{\rm col} + P_{\rm sp} &\propto N^2 G + \frac{\kr^2 N}{\pi},
\end{align}
where 
\begin{align}
	G &= \oint \frac{d^2 {\bf k}_\perp}{(2\pi)^2} |n_{\mathcal F}({\bf k}_0 - {\bf k}_\perp)|^2 \sim \frac{1}{A}
\end{align}
is a geometric factor and $A$ is the cross-sectional area of the system normal to the imaging axis (in detail this depends on the exact density distribution).
Thus the ratio of collective to spontaneous scattering is $\sim \pi N / (\kr^2 A)$ and the total number of spontaneously and collectively scattered photons is
\begin{align}
	N_{\rm sp} &= N\frac{g^2}{8}, & {\rm and} && N_{\rm col} &= \frac{\pi N^2 G}{\kr^2} \frac{g^2}{8}.
\end{align}
For our system with $N\approx 7\times10^4$ and transverse Thomas-Fermi radii $(R_x,R_y)\approx(43, 4)\ \mu{\rm m}$ the ratio of these is $\approx 6$. 

The integrated number of spontaneously scattered photons gives an SNR of $g \sqrt{N/8}$ for the determination of the atom number $N$. 
For example, for our BEC and $g=0.3$ about $N_{\rm sp}=800$ photons would be spontaneously scattered.
This implies that the forward scattered light would contain about $N_{\rm col}=5\times10^3$ photons.

In both cases the uncertainty in the photon number is simply the square root of the photon number.
The SNR in the atom number derived from spontaneously scattered light is therefore $\approx 30$ since the atom number is proportional to $N_{\rm sp}$.
For the forward scattered light, the atom number is proportional to $N_{\rm col}^{1/2}$ giving a SNR of $\approx 140$.
In both cases the SNR is proportional to $g$.

\section{R\lowercase{amsey Interferometry with Measurement Induced Decoherence}}\label{supp:RamseyInt}

We consider two-states $\left\{\ket{g_1}\equiv\ket{\downarrow}, \ket{g_2}\equiv\ket{\uparrow}\right\}$ that are microwave coupled with the rotating wave Hamiltonian
\begin{align*}
	\hat H^\prime(\phi_P) &= \frac{1}{2} \left[\Delta \hat \sigma_z + \Omega_\mu \sin(\phi_P) \hat \sigma_x - \Omega_\mu  \cos(\phi_P) \hat \sigma_y \right],
\end{align*}
in terms of the detuning $\Delta$ (in this case $\Delta >0$ corresponds to red detuning and $\Delta < 0$ yields blue detuning, this is reversed with respect to the spectroscopy convention), coupling strength $\Omega_\mu$ and microwave oscillator phase $\phi_P$.

\subsection{Standard Ramsey interferometer} 
We first outline the basic framework describing Ramsey interferometry (RI) for an arbitrary many-body state. 
Consider an atomic ensemble in an initial state $\ket{\Psi_i}$ along $-\ez$ on the Bloch sphere (with no population in $\ket{\uparrow}$), i.e., $\hat b_\uparrow \ket{\Psi_i} = 0$.

In the second quantized notation, the operator
\begin{align*}
	\hat R_z(\theta) & \equiv \exp\left[i\frac{\theta}{2}\hat b_i^\dagger \sigma_{z,ij} \hat b_j \right] = \exp\left[i\frac{\theta}{2}\left(\hat b_\uparrow^\dagger \hat b_\uparrow -  \hat b_\downarrow^\dagger \hat b_\downarrow\right)\right],
\end{align*}
implements a spin-rotation about $\ez$ by the angle $\theta$, and similarly for the other two axes (note that we used an implied summation convention in this expression).
Here $\sigma_{x,y,z}$ is the Pauli matrix for the specified axis. 
Our experimental sequence began with a $\pi/2$ pulse to rotate the system into the equal superposition state $(\ket{\uparrow} + \ket{\downarrow})/\sqrt{2}$, aligned along $\ex$ on the Bloch sphere.

The RI completed with a second $\pi/2$ pulse, that drove rotations about the $\ex \sin \delta\phi_{\rm P} -\ey \cos\delta\phi_{\rm P}$ axis (implemented by phase shifting the microwave oscillator by $\delta\phi_{\rm P}$). 
The final state is
\begin{equation}
	\ket{\Psi_f} = \left[\hat{R}_z^\dagger(\delta\phi_{\rm P})\hat{R}_y(\pi/2)\hat{R}_z(\delta\phi_{\rm P})\right]\hat{R}_y(\pi/2) \ket{\Psi_i}
	\label{Eqn:PsiF_Simple_RI},
\end{equation} 
where the quantity in square brackets implements the rotation about the new axis.
In this standard RI scheme, for the initial state with all atoms in $\ket{\downarrow}$ the number of atoms in state $\ket{\uparrow}$ in the final state is 
\begin{align}
	\langle\hat n_\uparrow\rangle &= \bra{\Psi_f} \hat b^\dagger_\uparrow \hat b_\uparrow \ket{\Psi_f} = N \cos^2 (\delta\phi_{\rm P}/2).
\end{align}
Here, the total number of atoms $N$ in the initial state is conserved in the RI process.

\subsection{Ramsey interferometer with a weak measurement}
In our experimental sequence with the dispersive-measurement pulse, the above classic RI is augmented with further evolution from the Kraus operator $\hat M({\bf k}_{\rm f})$ [see Eq.~(4) in the main text] describing the measurement.  
The final state is therefore
\begin{align*}
	\ket{\Psi_f} &= \hat{R}^\dagger_z(\delta\phi_{\rm P}) \hat{R}_y(\pi/2) \hat{R}_z(\delta\phi_{\rm P})\hat M({\bf k}_{\rm f})\hat{R}_y(\pi/2) \ket{\Psi_i}.
\end{align*} 

Our discussion for conventional RI was completely agnostic regarding the initial state.
This is not the case with the addition of a measurement pulse.
We describe our BEC as containing $N$ atoms in the same spatial mode $\psi({\bf x})$ [with Fourier transform $\tilde\psi({\bf k})$], and adopt the notation where
\begin{align}
	\hat b^\dagger &= \int d^3 {\bf x} \psi({\bf x}) \hat b^\dagger({\bf x}) =  \int  \frac{d^3 {\bf k}}{(2\pi)^3} \tilde\psi({\bf k}) \hat b^\dagger({\bf k})
\end{align}
describes the creation of a single particle in that mode (we will add $\uparrow\!/\!\downarrow$ subscripts as needed in what follows).
Accordingly our initial state is $\ket{N} = \left(\hat b^\dagger\right)^N \ket{0} / \sqrt{N!}$ and we will make frequent use of the relation $\hat b({\bf k}) \ket{N} = \sqrt{N}\tilde\psi({\bf k})\ket{N-1}$.

An alternate description in terms of coherent states is possible but makes non-physical predictions. 
In this description the initial state is an eigenstate of the annihilation operator with $\hat b \ket{\beta} = \beta \ket{\beta}$ where $N\rightarrow|\beta|^2$ is the average atom number.
In the measurement problem, this has the implication that each scattering processes creates a recoiling atom, {\it but} the number of particles in the condensate mode does not decrease since the action of $\hat b$ leaves $\ket{\beta}$ unchanged.

	We are interested in both the probability of detecting scattered photons and the change in RI contrast. 
	As before, we assume all atoms are initially in $\ket{\downarrow}$ number state.
	We begin with the scattering probability $P({\bf k}_{\rm f})$ giving
	\begin{align}
		P({\bf k}_\perp) &= \bra{N} \hat R^\dagger_{y}(\pi/2) \hat R^\dagger_z(\delta\phi_{\rm P}) \hat M^\dagger({\bf k}_\perp) R^\dagger_{y}(\pi/2) \hat R_{y}(\pi/2) \hat M({\bf k}_\perp) \hat R_z(\delta\phi_{\rm P}) \hat R_{y}(\pi/2) \ket{N} \nonumber \\ 
		&= \frac{t_{\rm m} P_{\rm e}}{c} |g({\bf k}_\perp)|^2 \frac{N}{2} \left[ \frac{N-1}{2} \left| \int \frac{d^3 {\bf k}_1}{(2\pi)^3} \tilde\psi^*_\downarrow({\bf k}_1) \tilde\psi_\downarrow[{\bf k}_1 - ({\bf k}_{\perp}-{\bf k}_0)]\right|^2 + 1 \right].
		\label{eq:P_scatter}
	\end{align}
	This reproduces Eq.~\eqref{eq:ScatProb} in the main manuscript but with atom number $N$ essentially reduced by 1/2 (owing to the $\pi/2$ pulse prior to applying the measurement pulse). 
	As elaborated on in the main text, the first term describes collective (stimulated) scattering with integrated probability $P_{\rm col}\propto N^2$ whereas the second term stems from spontaneous emission with probability $P_{\rm sp}\propto N$. 
	
	We are interested in the mean number of atoms in momentum state ${\bf k}$ and internal state $\ket{\uparrow}$, conditioned on detecting a photon in state ${\bf k}_\perp$:
	\begin{align}
		\langle\hat n_\uparrow({\bf k})\rangle_{|{\bf k}_\perp} &= \frac{1}{P({\bf k}_\perp)}\bra{N} \hat R^\dagger_{y}(\pi/2) \hat R^\dagger_z(\delta\phi_{\rm P}) \hat M^\dagger({\bf k}_\perp) R^\dagger_{y}(\pi/2) \hat b^\dagger_\uparrow({\bf k}) \hat b_\uparrow({\bf k}) \hat R_{y}(\pi/2) \hat M({\bf k}_\perp) \hat R_z(\delta\phi_{\rm P}) \hat R_{y}(\pi/2) \ket{N} \nonumber \\
		&= \frac{1}{2P({\bf k}_\perp)} \bra{N} \hat R^\dagger_{y}(\pi/2) \hat R^\dagger_z(\delta\phi_{\rm P}) \hat M^\dagger({\bf k}_\perp) \left[\hat b_\uparrow^\dagger({\bf k}) + \hat b_\downarrow^\dagger({\bf k})\right]\left[ \hat b_\uparrow({\bf k}) + \hat b_\downarrow({\bf k})\right] \hat M({\bf k}_\perp) \hat R_z(\delta\phi_{\rm P}) \hat R_{y}(\pi/2) \ket{N} \nonumber \\
		&\propto \frac{1}{2}\int \frac{d^3 {\bf k}_1}{(2\pi)^3} \frac{d^3 {\bf k}_2}{(2\pi)^3} \bra{N} \hat R^\dagger_{y}(\pi/2) \bigg\{\hat b_\uparrow^\dagger({\bf k}_1) \hat b_\uparrow[{\bf k}_1 - ({\bf k}_\perp-{\bf k}_0)] \left[e^{-i\delta\phi_{\rm P}/2}\hat b_\uparrow^\dagger({\bf k}) + e^{i\delta\phi_{\rm P}/2}\hat b_\downarrow^\dagger({\bf k})\right]  \nonumber\\
		& \ \ \ \times\left[ e^{i\delta\phi_{\rm P}/2}\hat b_\uparrow({\bf k}) + e^{-i\delta\phi_{\rm P}/2}\hat b_\downarrow({\bf k})\right] \hat b_\uparrow^\dagger[{\bf k}_2 - ({\bf k}_\perp-{\bf k}_0)] \hat b_\uparrow({\bf k}_2)\bigg\} \hat R_{y}(\pi/2) \ket{N}.\label{eq:contrast1}
	\end{align}
	In the last line, we omitted the numerical prefactor for brevity, but will reinstate it at the end of our computation.
	In the next step we normal order the field operators (with all creation operators moved to the left and the annihilation operators to the right), by performing commutators term-by-term on the chain of field operators inside the curly braces in Eq.~\eqref{eq:contrast1} (without the leading and trailing rotation operators).
	This procedure yields an expression
	\begin{align*}
		&= \hat b_\uparrow^\dagger({\bf k}_1) \left[e^{-i\delta\phi_{\rm P}/2}\hat b_\uparrow^\dagger({\bf k}) + e^{i\delta\phi_{\rm P}/2}\hat b_\downarrow^\dagger({\bf k})\right]\hat b_\uparrow^\dagger[{\bf k}_2 - ({\bf k}_{\perp}-{\bf k}_0)] \\
		&\ \ \ \ \ \times \hat b_\uparrow[{\bf k}_1 - ({\bf k}_{\perp}-{\bf k}_0)]  \left[ e^{i\delta\phi_{\rm P}/2}\hat b_\uparrow({\bf k}) + e^{-i\delta\phi_{\rm P}/2}\hat b_\downarrow({\bf k})\right] \hat b_\uparrow({\bf k}_2) \\
		& + \delta({\bf k}_1 - {\bf k}_2)\hat b_\uparrow^\dagger({\bf k}_1) \left[e^{-i\delta\phi_{\rm P}/2}\hat b_\uparrow^\dagger({\bf k}) + e^{i\delta\phi_{\rm P}/2}\hat b_\downarrow^\dagger({\bf k})\right] \left[ e^{i\delta\phi_{\rm P}/2}\hat b_\uparrow({\bf k}) + e^{-i\delta\phi_{\rm P}/2}\hat b_\downarrow({\bf k})\right] \hat b_\uparrow({\bf k}_2) \\
		& + \delta({\bf k}_1 - ({\bf k}_{\perp}-{\bf k}_0) - {\bf k}) e^{-i\delta\phi_{\rm P}/2}\hat b_\uparrow^\dagger({\bf k}_1) \hat b_\uparrow^\dagger[{\bf k}_2 - ({\bf k}_{\perp}-{\bf k}_0)] \left[ e^{i\delta\phi_{\rm P}/2}\hat b_\uparrow({\bf k}) + e^{-i\delta\phi_{\rm P}/2}\hat b_\downarrow({\bf k})\right] \hat b_\uparrow({\bf k}_2) \\
		& + \delta({\bf k}_2 - ({\bf k}_{\perp}-{\bf k}_0) - {\bf k}) e^{i\delta\phi_{\rm P}/2} \hat b_\uparrow^\dagger({\bf k}_1) \left[e^{-i\delta\phi_{\rm P}/2}\hat b_\uparrow^\dagger({\bf k}) + e^{i\delta\phi_{\rm P}/2}\hat b_\downarrow^\dagger({\bf k})\right]  \hat b_\uparrow[{\bf k}_1 - ({\bf k}_{\perp}-{\bf k}_0)] \hat b_\uparrow({\bf k}_2) \\
		& + \delta({\bf k}_1 - ({\bf k}_{\perp}-{\bf k}_0) - {\bf k}) \delta({\bf k}_2 - ({\bf k}_{\perp}-{\bf k}_0) - {\bf k}) \hat b^\dagger_\uparrow({\bf k}_1) \hat b_\uparrow({\bf k}_2),
	\end{align*}
	with five terms that we label $({\rm I})$ to $({\rm V})$ from top to bottom.
	We evaluate each term (now accounting for the rotation operators) by keeping only the operators that act on the initial $\ket{\downarrow}$ state.
	Integrating over all BEC momentum states ${\bf k}$, we find
	\begin{align}
		({\rm I}) &= \frac{N(N-1)(N-2)}{4} \cos^2 \left(\frac{\delta\phi_{\rm P}}{2}\right) \left| \int \frac{d^3 {\bf k}_1}{(2\pi)^3} \tilde\psi^*_\downarrow({\bf k}_1) \tilde\psi_\downarrow[{\bf k}_1 - ({\bf k}_{\perp}-{\bf k}_0)]\right|^2, \\
		({\rm II}) &= \frac{N(N-1)}{2} \cos^2 \left(\frac{\delta\phi_{\rm P}}{2}\right),\\
		({\rm III}) + ({\rm IV}) &=  \frac{N(N-1)}{2} \cos^2 \left(\frac{\delta\phi_{\rm P}}{2}\right) \left| \int \frac{d^3 {\bf k}_2}{(2\pi)^3} \tilde\psi_\downarrow^*[{\bf k}_2 - ({\bf k}_{\perp}-{\bf k}_0)]\tilde\psi_\downarrow({\bf k}_2)\right|^2,\\
		({\rm V}) & = \frac{N}{4}.
	\end{align}
Together terms (I), (III) and (IV) describe collective scattering (the usual Born and Wolf~[S1] forward scattering from the overall density distribution, as previously noted in the main manuscript) and combine to give  
	\begin{align}
		({\rm I}) + ({\rm III}) + ({\rm IV}) &= \frac{N(N-1)N}{4} \cos^2 \left(\frac{\delta\phi_{\rm P}}{2}\right) \left| \int \frac{d^3 {\bf k}_1}{(2\pi)^3} \tilde\psi^*_\downarrow({\bf k}_1) \tilde\psi_\downarrow[{\bf k}_1 - ({\bf k}_{\perp}-{\bf k}_0)]\right|^2.
	\end{align}
	On the other hand, terms (II) and (V) give
	\begin{align}
		({\rm II}) + ({\rm V}) &= \frac{N}{2}\left[(N-1) \cos^2 \left(\frac{\delta\phi_{\rm P}}{2}\right) + \frac{1}{2} \right],
	\end{align}
	describing independent scattering, i.e., spontaneous emission. 
	These two terms convey that the contrast will be reduced by the fractional count of an atom, i.e. $1/N$, for each spontaneous scattering event.

	Finally, we combine all five terms, re-insert the prefactor, integrate over ${\bf k}_\perp$ and organize into contributions from single atom scattering  $P_{\rm sp}$ and collective scattering $P_{\rm col}$.
	This elucidates the distinction between large angle and small angle scattering, giving
	\begin{align}
		\langle\hat N_\uparrow(\delta\phi_{\rm P})\rangle  &=  \frac{P_{\rm col}}{P_{\rm tot}} N \cos^2 \left(\frac{\delta\phi_{\rm P}}{2}\right) + \frac{P_{\rm sp}}{P_{\rm tot}} \left[\left(N-1\right)\cos^2 \left(\frac{\delta\phi_{\rm P}}{2}\right)+\frac{1}{2}\right],
	\end{align}
with total scattering probability $P_{\rm tot}$.
In the first term, small angle collective scattering returns atoms into their initial spatial mode (Mössbauer scattering); thereby leaving the RI contrast unchanged.
In the second term, larger angle scattering sends the recoiling atom into a previously empty mode.
In this process the environment counts the scattered atom to be in $\ket{\uparrow}$ and removes its contribution to the contrast, as motivated in the main manuscript. 

\section{L\lowercase{attice pulse sequence}}\label{supp:Lattice}

\begin{figure*}[htb]
	\includegraphics{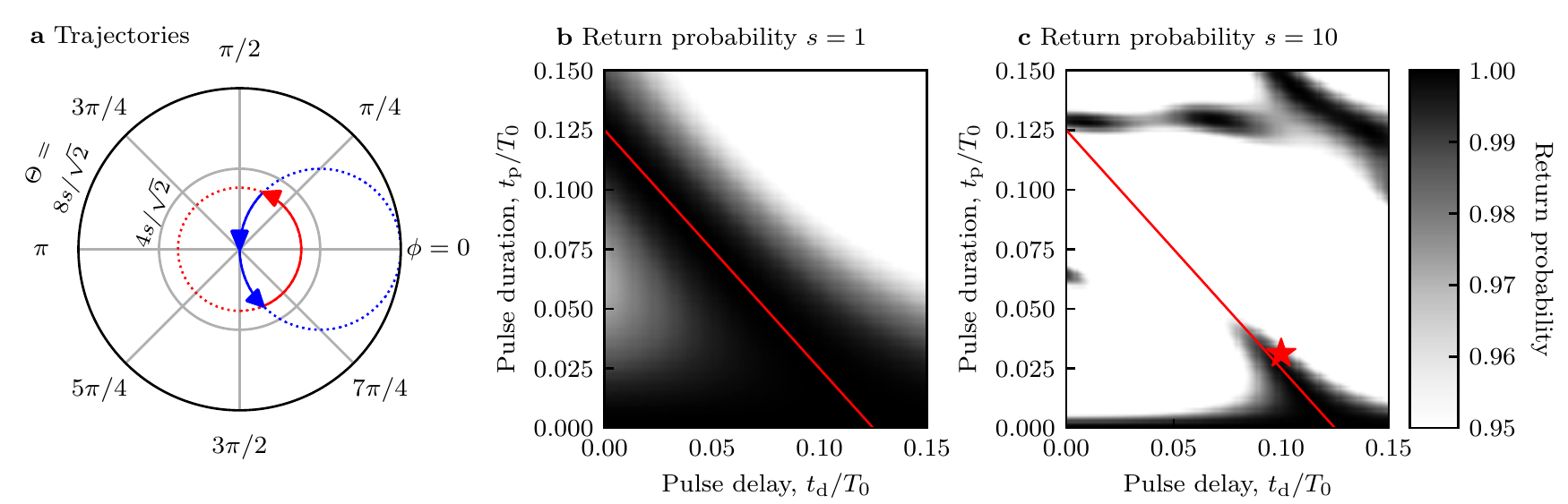}
	\caption{
		Lattice pulse sequence.
		{\bfseries a} Top of Bloch sphere for the spherical polar angles $\phi$ and $\theta$ showing trajectories for zero (red) and non-zero (blue) $s$.
		The dashed curve shows complete orbits while the solid curve with arrows results from a pulsed sequence that combine to form trajectory that returns to the initial state.
		{\bfseries b} and {\bfseries c} Return probability computed including $7$ momentum states for $s=1$ and $s=10$.
		The star marks the parameters used in our experiment.
	}
	\label{Fig:SMLattice}
\end{figure*}

This section shows the diagrams elucidating the lattice pulse sequence used to remedy the effect of the stray weak optical lattice. 
See the second subsection of the Method section in the main manuscript for a more detailed analysis.

\section{M\lowercase{agnetic field lock}}
\label{Supp:uwaveLock}

\begin{figure}[htb]
	\includegraphics{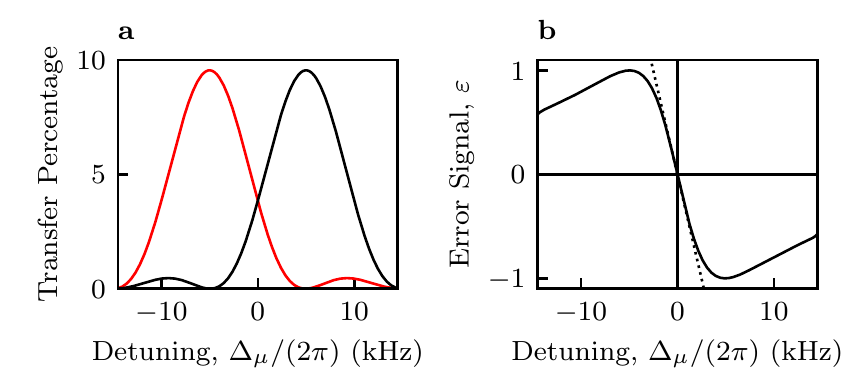}
	\caption{
		Microwave field lock.
		{\bfseries a} Transfer fraction for individual pulses with parameters as described in subsection A of the Method section in the main manuscript.
		{\bfseries b} Error function (solid curve) and linear approximation (dashed line).
	}
	\label{Fig:FieldLock}
\end{figure}

This section presents the diagrams for the microwave field lock employed to minimize the impact of magnetic field shifts in Ramsey contrast measurements. 
See the first subsection of the main manuscript Method section for a more detailed description of this procedure.

\section*{S\lowercase{upplementary} R\lowercase{eferences}}

[S1] M. Born and E. Wolf, {\it Principles of Optics: Electromagnetic Theory of Propagation, Interference and Diffraction of Light} (7th Edition), 7th ed. (Cambridge University Press, 1999).

\end{document}